\documentclass[preprint,singlecolumn]{revtex4-1}
\usepackage{graphicx}
\usepackage{natbib}
\usepackage{xcolor}
\usepackage{tikz}
\usepackage{subfigure}
\usepackage{epstopdf}
\usepackage{booktabs}
\usepackage{amsmath}
\usepackage{amssymb}
\usepackage{wasysym}
\usepackage{enumerate}
\usepackage{pgf}

\begin{document}

%%%%%%%%%%%%%%% ABSTRACT %%%%%%%%%%%%%%%%%%%%%
\begin{abstract}
A scaling relation for the high frequency regime of streamwise energy spectra (in frequency domain) in the near-wall region is proposed. This is based on the dimensional analysis approach of \cite{perry1977} and \cite{Zamalloa2014} together with the hypothesis that the small-scale fluctuations in the near-wall region should only depend on the viscous scales, analogous to the Prandtl's law-of-the-wall for the mean flow. This allows us to examine the lower bound for the high frequency regime where law-of-the-wall in spectra would hold. Observations in high-Reynolds-number  turbulent boundary layer data indicate that a conservative estimate for the start of this high frequency regime is $f^+$ = 0.005 (which corresponds to 200 viscous time-units) across a range of wall-normal positions and Reynolds numbers.  This is sufficient to capture the energetic viscous-scaled motions such as the near-wall streaks, which has a time scale of approximately 100 viscous units. This scaling relation and the spectral collapse is consistent with the observations in internal flows at lower Reynolds numbers \cite{Zamalloa2014}.  
\end{abstract}

% Use the \preprint command to place your local institutional report
% number in the upper righthand corner of the title page in preprint mode.
% Multiple \preprint commands are allowed.
% Use the 'preprintnumbers' class option to override journal defaults
% to display numbers if necessary
%\preprint{}

%Title of paper
\title{Law-of-the-wall for streamwise energy spectra in \\high-Reynolds-number turbulent boundary layers}

% repeat the \author .. \affiliation  etc. as needed
% \email, \thanks, \homepage, \altaffiliation all apply to the current
% author. Explanatory text should go in the []'s, actual e-mail
% address or url should go in the {}'s for \email and \homepage.
% Please use the appropriate macro foreach each type of information

% \affiliation command applies to all authors since the last
% \affiliation command. The \affiliation command should follow the
% other information
% \affiliation can be followed by \email, \homepage, \thanks as well.
\author{B. Ganapathisubramani} 
\email[]{g.bharath@soton.ac.uk}
%\homepage[]{Your web page}
%\thanks{}
%\altaffiliation{}
\affiliation{\small University of Southampton, Southampton SO17 1BJ, UK }

%Collaboration name if desired (requires use of superscriptaddress
%option in \documentclass). \noaffiliation is required (may also be
%used with the \author command).
%\collaboration can be followed by \email, \homepage, \thanks as well.
%\collaboration{}
%\noaffiliation

\date{\today}

\maketitle

\section{Background}

A number of previous efforts have focussed on proposing scaling laws for turbulent energy spectra of turbulent wall-flows. Perry \& Abell \cite{perry1977} was perhaps the first to develop scaling laws for streamwise energy spectra using dimensional analysis approach. This work has been extensively extended in various subsequent studies \citep{Perry1982,Perry1986,Perry1995}. The primary focus of these efforts were on the overlap scaling of streamwise energy spectra where the energy content is inversely proportional to the wall-normal position. The choice of different velocity and length-scales for the dimensional analysis in these previous studies were underpinned by the attached-eddy hypothesis, a model of wall turbulence that provides critical insights. Very recently, Zamalloa {\it et al.}\cite {Zamalloa2014} used dimensional analysis and proposed ``model-free'' scaling relations for the turbulent-energy spectra in different regions (near-wall, log and outer wake regions) of turbulent wall-flows. Specifically, they demonstrated the presence of law-of-the-wall in the high wavenumber regime of energy spectra in pipe/channel flows using experimental and DNS data over a limited range of Reynolds numbers. 

In this paper, we present a re-interpretation of the dimensional analysis work carried out by Perry \& Abell \cite{perry1977} and Zamalloa {\it et al.} \cite{Zamalloa2014} as well as observations  of law-of-the-wall in small-scale turbulence intensity and in the high frequency regime of streamwise energy spectra in high Reynolds number turbulent boundary layers. We highlight some subtle differences (compared to those previous studies) in the choice of velocity and length scales (used in dimensional analysis) and justify these choices based on physical reasoning and experimental observations. 

\section{Law-of-the-wall for streamwise energy spectra}

This section is a recap of the dimensional analysis performed by \cite{perry1977} and \cite{Zamalloa2014}. However, we highlight the choice of different scaling variables and the reasoning required for it. Let $\phi_{11}(k_1,y)$  be the power spectral density of the streamwise velocity fluctuation for a longitudinal wavenumber $k_1$ at location $y$ away from the wall. Then, the integral over all $k_1$ of this power spectral density is equal to the turbulent energy of the streamwise velocity component at that wall-normal location. 

\begin{equation}
\int_0^\infty \phi_{11}(k_1,y) dk_1 = u^2(y)
\end{equation}

Now, $k_1\phi_{11}(k_1,y)$ is the turbulent energy contained at a given wavenumber $k_1$ at a certain wall-normal location. Following \cite{perry1977} and \cite{Zamalloa2014} and their dimensional analysis approach, it is clear that we need a velocity scale to non-dimensionalise $k_1\phi_{11}$, and potentially two different length scales for the wavenumber and wall-normal position. 

\begin{equation}
\frac{k_1\phi_{11}(k_1,y)}{\tilde{U}^2} = F(kL_1,y/L_2)
\end{equation}

\noindent
where, $L_1$ is the relevant length scale for wavenumber and $L_2$ is the relevant length-scale for wall-normal position. $\tilde{U}$ is the relevant velocity scale for the energy. We have to to use physical arguments to arrive at appropriate choices for these scales.

Recall that Prandtl's law-of-the-wall for mean flow postulates that at high Reynolds numbers, close to the wall ($y<<\delta$, where $\delta$ is the outer length scale), there is an inner layer in which the mean velocity is determined by the viscous scales, independent of outer length and velocity scales. Based on this, we arrive at an equation for the mean-flow, which is the law-of-the-wall:

\begin{equation}
\frac{U}{U_\tau} = f_w\left(\frac{yU_\tau}{\nu}\right)
\end{equation}

$U_\tau$ is the skin-friction velocity ($U_\tau = \sqrt{\tau_w/\rho}$, where $\tau_w$ is the wall-shear-stress and $\rho$ is the density of the fluid) and $\nu$ is the kinematic viscosity of the fluid.  

We can follow the same reasoning as for the mean profile to determine the appropriate values for $\tilde{U}$, $L_1$ and $L_2$. As an observer at the wall with ``no knowledge'' of the outer flow, there is only one choice for $\tilde{U}$, which is the viscous velocity scale ($U_\tau$).  There may be two possible candidates for $L_1$: viscous length scale ($\nu/U_\tau$) and wall-normal position ($y$). For $L_2$, there is only one possible candidate, which is the viscous length scale $\nu/U_\tau$ (note that  we will represent scales non-dimensionalised with the viscous scales with a superscript `+'). Then, the spectral scaling becomes, 
 
\begin{equation}
\frac{k_1\phi_{11}(k_1,y)}{U^2_\tau} = F(k_1L_1,y^+)
\end{equation}

%Note that this analysis is identical to the one in \cite{Perry1977} where the authors postulated a universal wall-scaling by using $L_1$ = $y$. They also resorted to Kolmogorov scaling at high wavenumbers following similar reasoning. They used $u_\eta$ and $\eta$ for the velocity and length scales in the high-wave number regime. However, we recognise that in the near-wall region, the viscous scaling from shear-stress at the wall is a more stringent condition compared to the Kolmogrov scales.  \cite{Perry1977} could just as easily resorted to inner-scaling and would have arrived at this result. 

%\begin{equation}
%\frac{\phi_{11}(k_1)}{U^3_\tau/\nu} = F(k^+_1)
%\end{equation}

\cite{perry1977} used their insights from attached-eddy hypothesis and chose $L_1 = y$ that highlights distance from the wall scaling.  They used $U_\tau$ to be the velocity scale. Here, we take a different approach to determine $L_1$. We postulate that there exists a law-of-the-wall for the small-scale velocity fluctuations, i.e. the integral of the spectrum over a certain high wavenumber range. In the near-wall region, the variance of the small-scales over a range of high wavenumbers is only influenced by inner velocity scale such that, 

\begin{equation}
\frac{\overline{u^2_S}}{U^2_\tau} = g_w(y^+)
\end{equation}

where, $u^2_S$ is the variance of the streamwise velocity fluctuations in the high wavenumber regime that should conform to law-of-the-wall and $g_w$ is the function that represents law-of-the-wall for these small-scale motions. This will result in a universal form that is only dependent on viscous scales and is independent of outer influence. Therefore, we can choose $L_1$ as $\nu/U_\tau$. For these high wavenumbers, \cite{perry1977} used the classical Kolmogorov scaling of $\eta$ (length scale) and $u_\eta$ (velocity scale) that depends on the dissipation of turbulent kinetic energy. However, we recognise that in the near-wall region, the inner-scale is a much more stringent requirement compared to Kolmogorov scale (which varies with wall-normal direction).  Based on this choice of $L_1$, we seek collapse in spectra such that,

\begin{equation}
g_w(y^+)  = \int_{M^+}^\infty  F(k^+_1,y^+) d[ln(k^+_1)]
\end{equation}

is a universal function in the inner region across Reynolds numbers. At fixed values of $y^+$ close enough to the wall (i.e. $y <<\delta$), the above equation indicates that there  should be a universal value of $M^+$ that is independent of $y^+$. 

Zamalloa {\it et al.} \cite{Zamalloa2014} use $y$ as their length scale $L_1$, however, they recognised that this scaling can be replaced with a scaling similar to the one mentioned above. They did not examine this in any detail. They used data and observed that the spectra collapses for high wavenumbers, if, $k_1 y$ $\geq$ $10(y^+/Re_\tau)$. This translates to $k^+_1$ $\geq$ $10/Re_\tau$ or $k_1\delta \geq 10$, which essentially means that there should be one order of magnitude difference in the wavenumber range compared to the large-scales of the flow. These observations were limited in Reynolds number range  (only up to $Re_\tau \approx 3000$) and were confined to internal flows. This limit of collapse might have an outer influence since the scale separation is not sufficiently large. Therefore, it is important to examine this collapse at higher Reynolds numbers to determine if there exists an appropriate wavenumber cut-off that eliminates the dependence of distance from the wall (or Reynolds number). This is examined in more detail in the next section.

%
%However, given that there is an outer influence, an appropriate correction to the above equation would be, 
%
%\begin{equation}
%\frac{\overline{u^2_S}(y^+)}{U^2_\tau}  = \int_{M^+(y^+)}^\infty  F(k^+_1,y^+) d[ln(k^+_1)]
%\end{equation}

%This equation indicates that in fact, it might be possible to choose the length scale $L_1$ as $y$ (wall-normal location). Therefore, the spectral scaling can be re-written as, 
%
%\begin{equation}
%\frac{k_1\phi_{11}(k_1,y)}{U^2_\tau} = F(k_1y,y^+)
%\end{equation}
%
%In turn, the law-of-the-wall for the energy contained within a range of high wavenumbers would become, 
%
%\begin{equation}
%\frac{u^2_S(y^+)}{U^2_\tau}  = \int_{M}^\infty  F(k_1y,y^+) d[ln(k_1y)]
%\end{equation}
%
%where, $M$ is a fixed value of $k_1y$ at matched $y^+$. 

\section{Observations of law-of-the-wall in frequency domain}

It is very difficult to obtain wavenumber spectra at high Reynolds number without invoking Taylor's hypothesis where we have to assume or model a frequency-wavenumber mapping function. Typically, this mapping function is assumed to just depend on the local mean velocity. This could lead to incorrect observations on the nature of collapse of spectra. Therefore, here we resort to using frequency spectra from hot-wire data without invoking Taylor's hypothesis (or other equivalent mapping). In this case, law-of-the-wall for spectra becomes, 

\begin{equation}
\frac{f\phi_{11}(f,y)}{U^2_\tau} = F(f T_1,y^+)
\end{equation}

where, $f$ is the frequency and $T_1$ is a suitable time-scale. Following the same arguments as in previous section, the appropriate time-scale should be the inner time-scale based on wall-shear stress and viscosity, $T_1 = \nu/U^2_\tau$. This makes the law-of-the-wall for streamwise energy spectrum in frequency domain, 

\begin{equation}
\frac{f\phi_{11}(f,y)}{U^2_\tau} = F(f^+,y^+)
\end{equation}

We now seek collapse in spectra across Reynolds numbers such that, 

\begin{equation} 
g_w(y^+) = \frac{u^2_S(y^+)}{U^2_\tau}  = \int_{A^+}^\infty  F(f^+,y^+) d[ln(f^+)]\label{eqnus}
\end{equation}
 is a universal function in the inner region. Here, $A^+$ is equivalent to $M^+$ where at fixed values of $y^+$ close enough to the wall (i.e. $y <<\delta$), there is a universal value of $A^+$.

\begin{figure}
\vspace{0.1in}
\begin{center}
\includegraphics[width=0.32\textwidth]{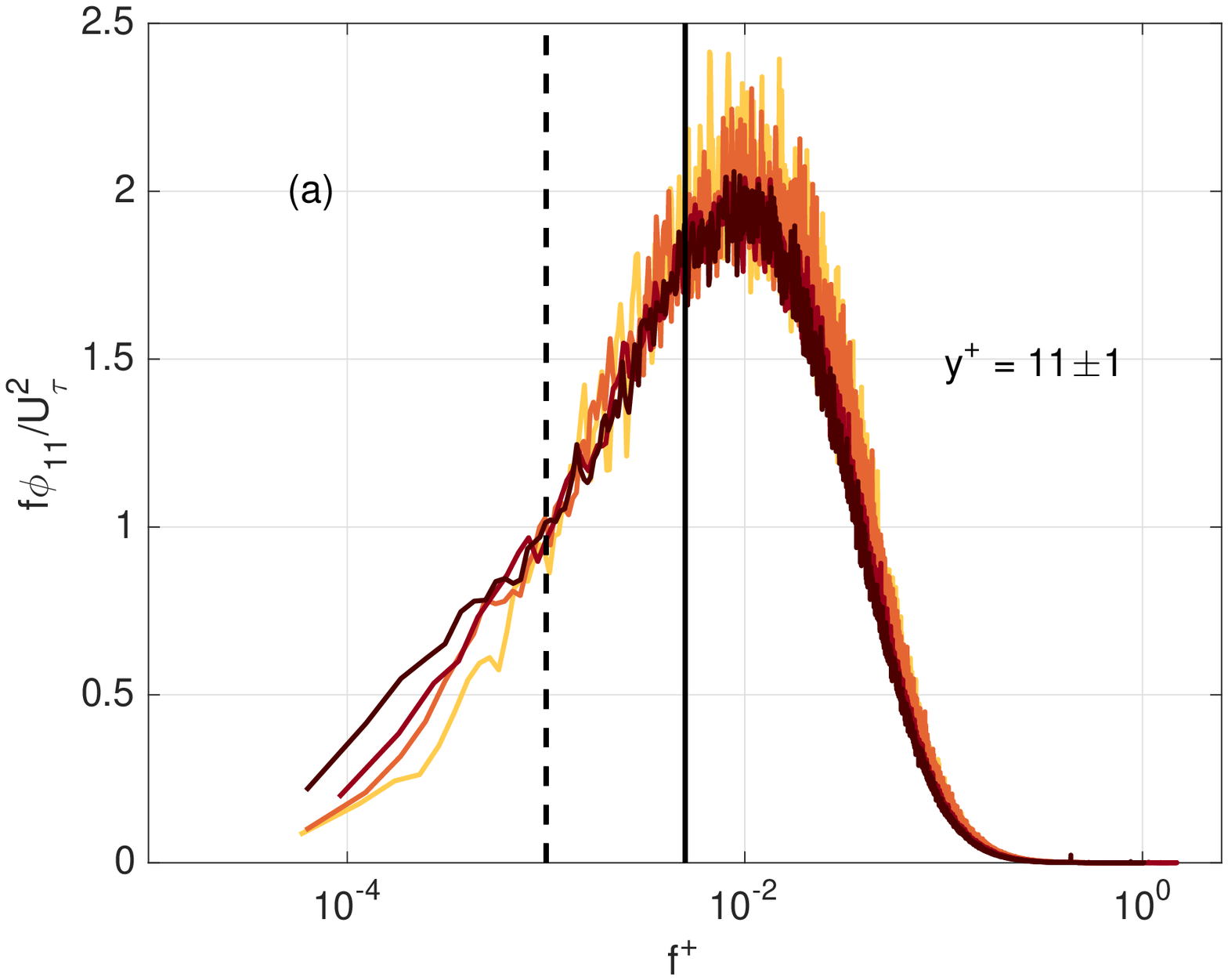}
\includegraphics[width=0.32\textwidth]{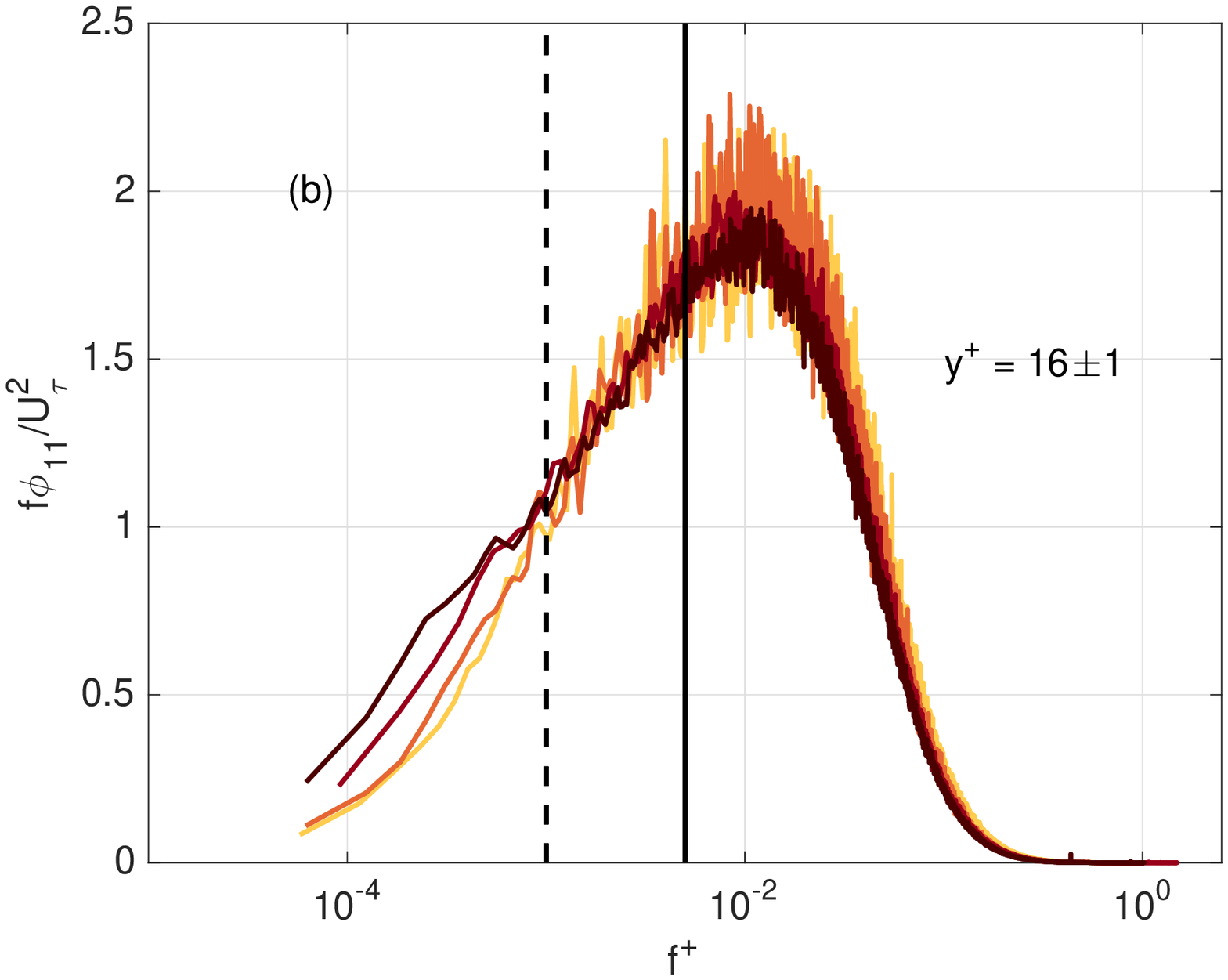}
\includegraphics[width=0.32\textwidth]{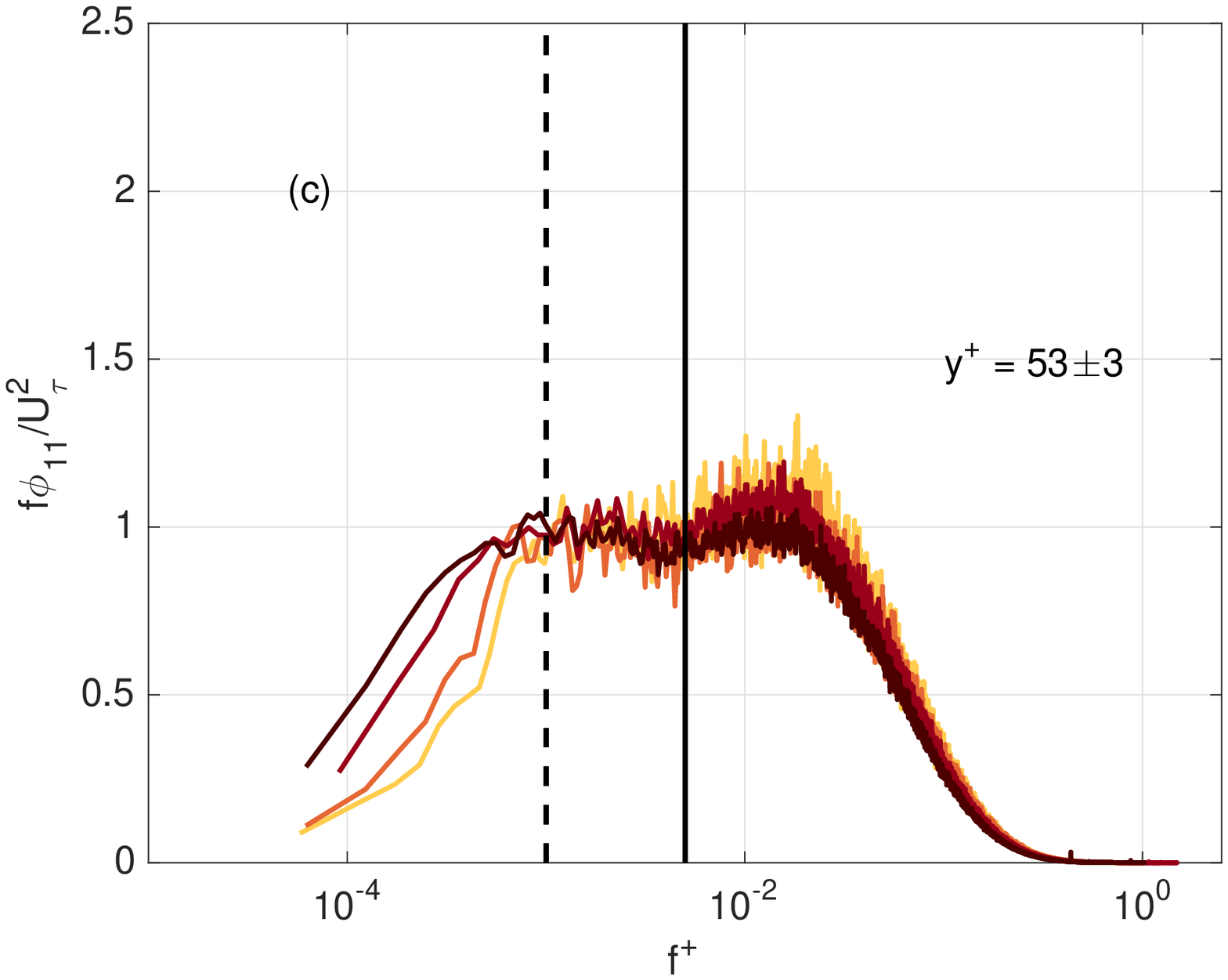}
\includegraphics[width=0.32\textwidth]{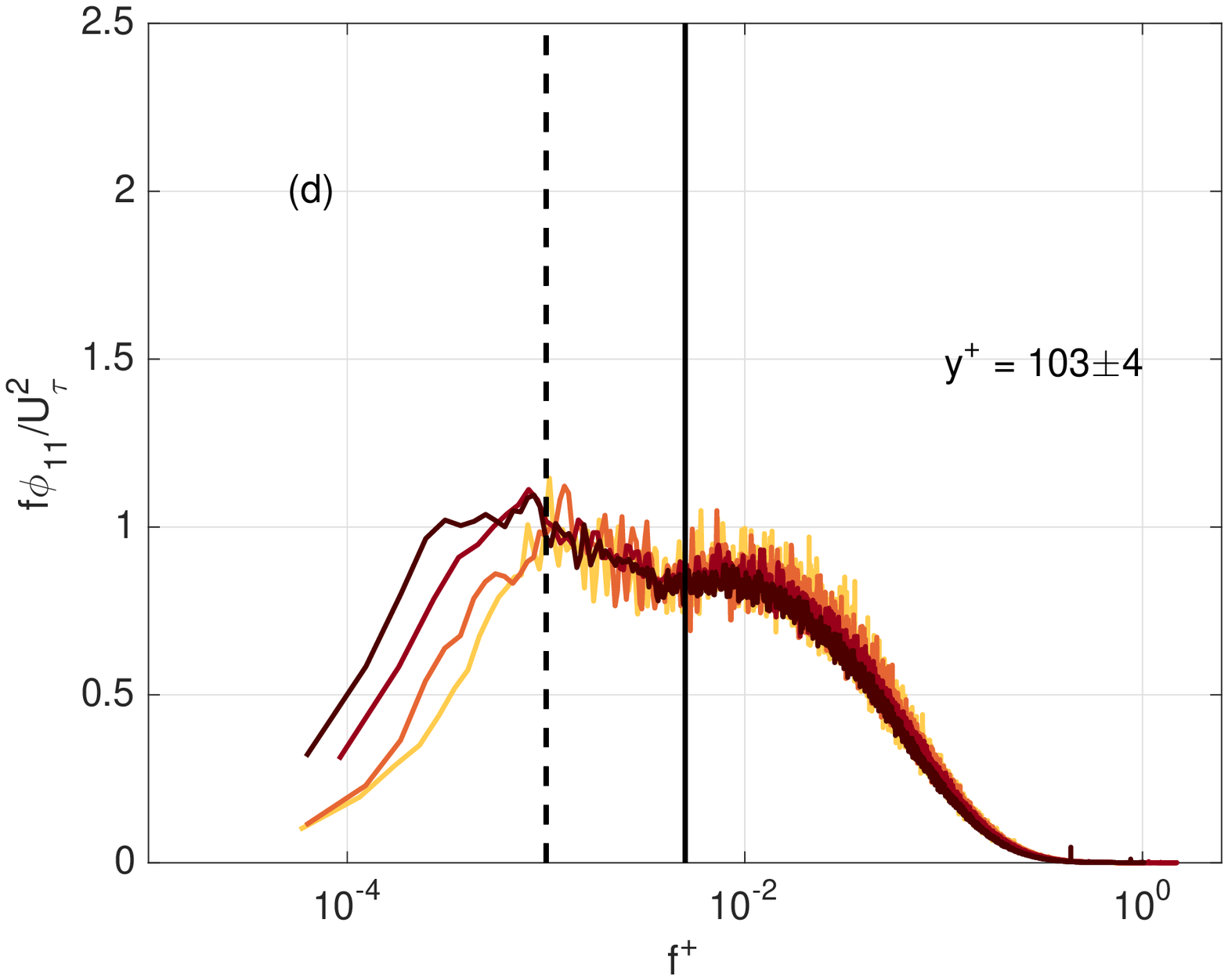}
\includegraphics[width=0.32\textwidth]{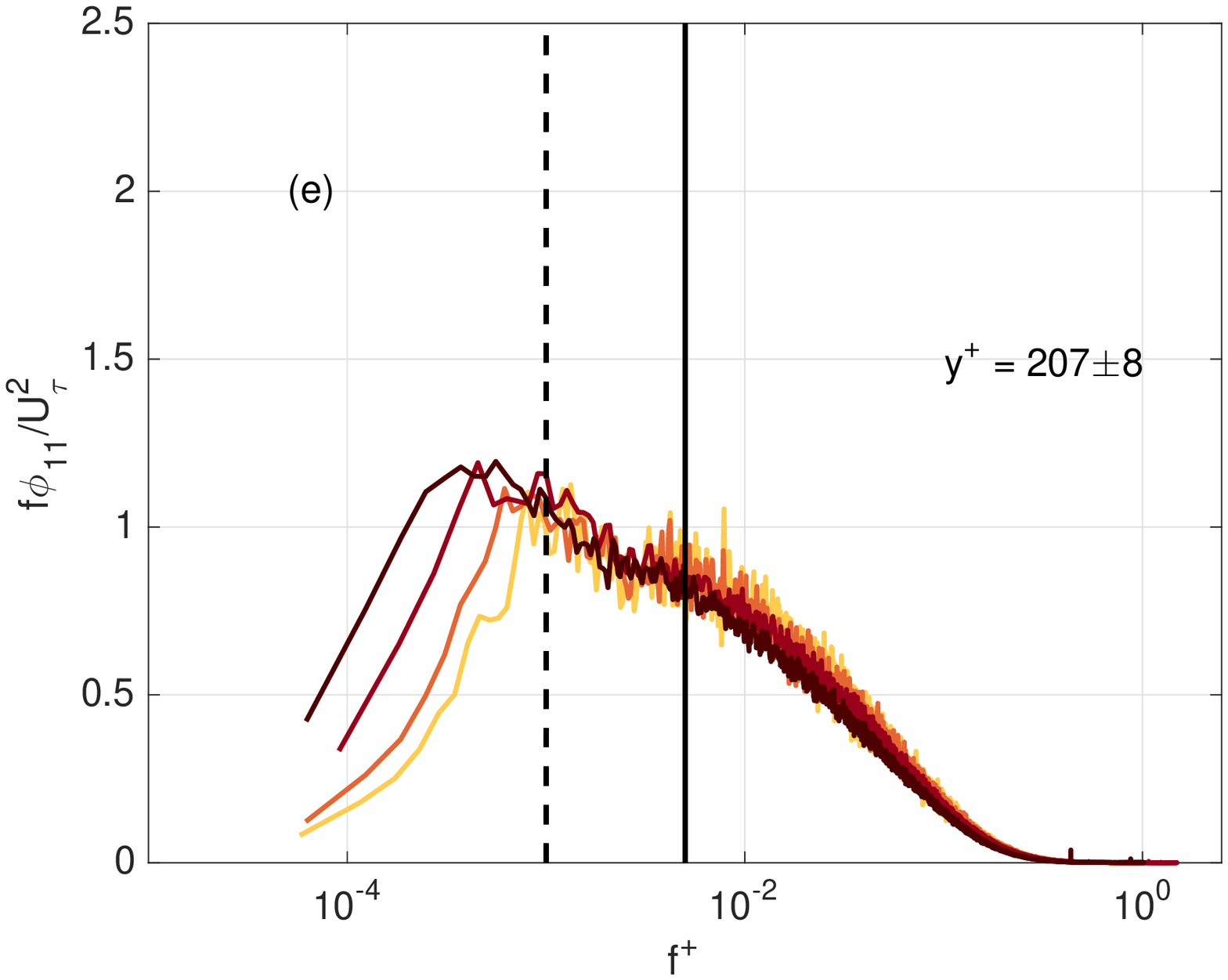}
\includegraphics[width=0.32\textwidth]{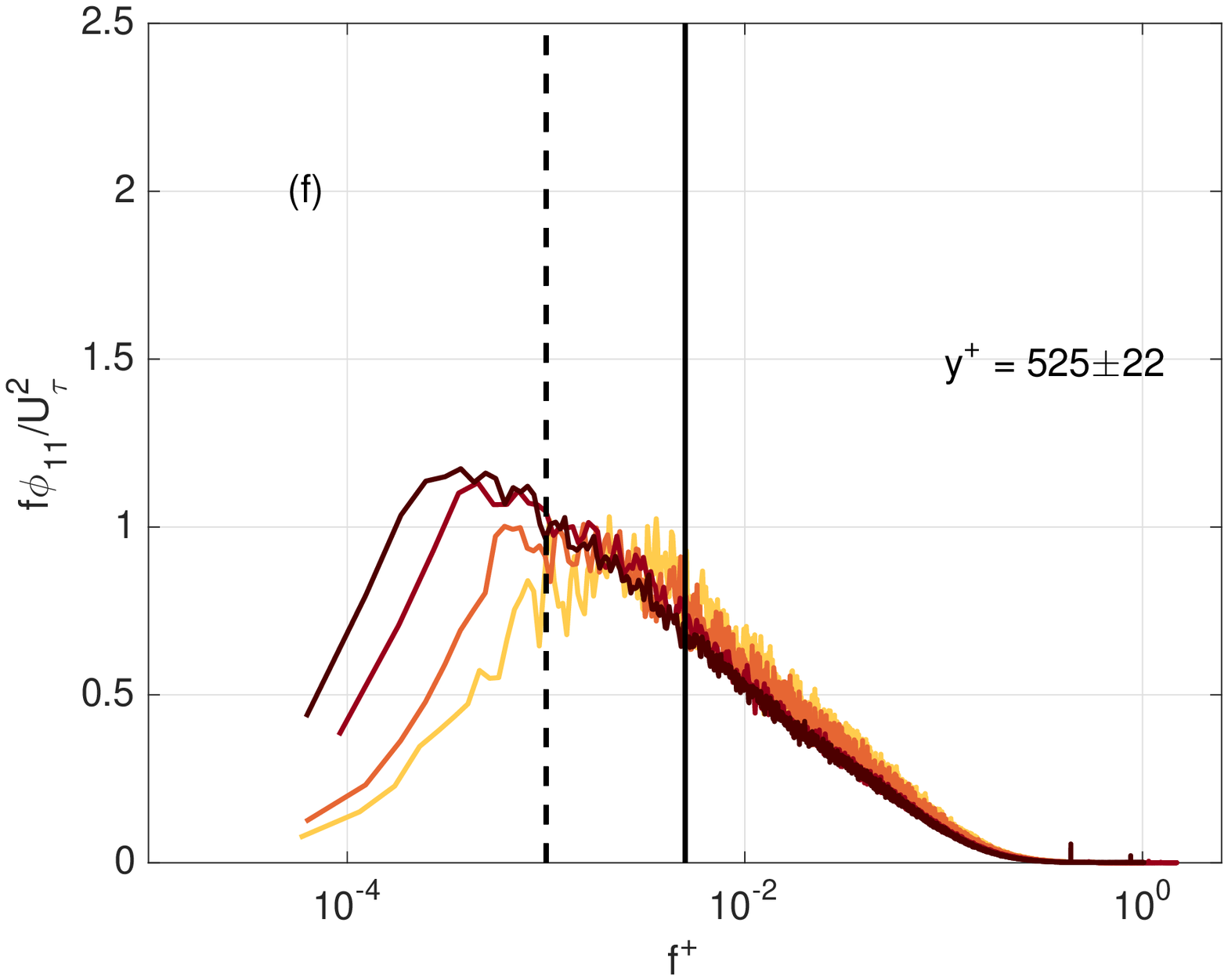}
\caption{($a)-(f)$ Inner-normalised energy spectra for 4 different Reynolds numbers at 6 different values of $y^+$. The figures show the value of $y^+$ and the variance in that selected wall-normal location across the Reynolds numbers. Lines are coloured from lightest to darkest in order of increasing $Re_\tau$ (= 2800, 3900, 7300 and 14150). The solid black line shows the $f^+$ = 0.005, which is 200 inner time-units while the dashed black line shows $f^+$ = 0.001, which is 1000 inner time-units.}
\label{fig1}
\end{center}
\end{figure}

Hot-wire measurements obtained in Melbourne's High Reynolds Number Boundary Layer Wind Tunnel (HRNBLWT) is used to compare spectra at similar wall-normal locations over a range of Reynolds numbers ($Re_\tau$ = 2800, 3900, 7300, 14150). The datasets for $Re_\tau$ = 2800, 3900 and 7300 are obtained from \cite{Hutchins2009} and \cite{Mathis2009}. The dataset for $Re_\tau$ = 14150 is from \cite{Hutchins2011}. Figure \ref{fig1} shows the energy spectra against inner-normalised frequency over a range of wall-normal locations including $y^+$ = 11$\pm1,$ 16$\pm1$, 53$\pm3$, 103$\pm4$, 207$\pm8$ and $525\pm22$. The largest $y^+$ value is chosen to be in outer-edge of the log region at the lowest Reynolds number. It is difficult to obtain data at exactly the same $y^+$ values across different Reynolds numbers. Therefore, the locations mentioned are average of the nearest wall-normal location over the range of Reynolds numbers and the error is based on twice the standard deviation of the wall-normal positions that are chosen across Reynolds numbers. 

Spectra in figure \ref{fig1} shows excellent collapse in the spectra at high frequencies where we expect the law-of-the-wall to hold. It should be noted the data at the highest Reynolds number might suffer from mild attenuation of energy due to spatial resolution of the hot-wire probe, especially, at these high frequencies \cite{Hutchins2009}. Despite this, the collapse is within the statistical error. The collapse is obvious even at $y^+ \approx 103$ down to a frequency of $f^+ = 0.001$ (which is shown in dashed line in the figure).  Below this frequency, the outer-influence becomes obvious and the spectrum for each Reynolds number peel off at different frequencies.  The quality of the collapse is worse at $y^+ = 525$. At that location, wall-normal location is at the outer-edge of the log region at $Re_\tau$ = 2800, while it is still in the log region for the other Reynolds numbers. We expect the law-of-the-wall for the small-scales of the spectrum to diminish at these locations where the shear is minimal and therefore the spectra at the small-scales would collapse with Kolmogorov scales. Regardless, in the near-wall region (where $y^+$ is small and $y<<\delta$), there is excellent collapse in the spectra  at least up to $f^+ = 0.005$ (shown as solid black line) and perhaps up to $f^+ = 0.001$ (dashed line).

\begin{figure}
\vspace{0.1in}
\begin{center}
\includegraphics[width=0.46\textwidth]{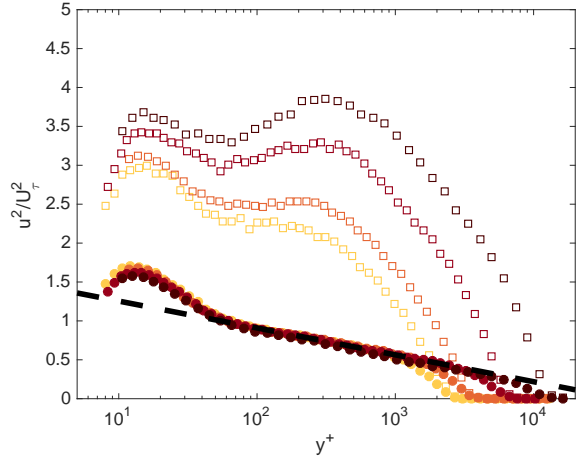}
\caption{Inner normalised large-scale (open squares) and small-scale (filled circles) variance of streamwise velocity for 4 different Reynolds numbers (increasing darkness represents increasing Reynolds number). The separation between large- and small-scale is based on a Fourier filter at $f^+$ = 0.005. The dashed black line shows the log decay of the small-scale variance beyond $y^+ > 100$.}
\label{fig2}
\end{center}
\end{figure}

Based on the above observations, a conservative estimate for $A^+$ in equation \ref{eqnus} can be 0.005.  We can now integrate the power spectral density from this value of $A^+$ up to infinity. This should give the near-wall small-scale turbulent energy, which obeys law-of-the-wall. The amount of energy contained in these scales will diminish farther away from the wall. In fact, farther away from the wall, the viscous scale may no longer be equal to inner-scales but could be replaced with Kolmogorov scales as in \cite{perry1977}. 

The impact of using $A^+  = 0.005$ as the cut-off frequency is explored by computing the large-scale and small-scale variance of streamwise velocity based on this cut-off value. Figure \ref{fig2} shows these two quantities in inner-scaling. The small-scale fluctuations across Reynolds numbers collapse across the entire range of wall-normal locations (up to $y^+ \approx$ 1000, which is well outside the log region for the lower Reynolds number).  In fact, the small-scale variance appears to follow a logarithmic decay with wall-normal position for $y^+ > 100$ across all Reynolds numbers. This logarithmic decay extends well in to the outer region (beyond the traditional log region for the mean profile). This log decay has the form, 

\begin{equation}
 \frac{u^2_S(y^+)}{U^2_\tau} = -0.15~{\rm ln}(y^+) + 1.6~{\rm for}~y^+ > 100
\end{equation}

The implications of the two constants (slope = -0.15 and intercept = 1.6) is unclear at this time and requires further interpretation. However, it does show that there is indeed a universal law-of-the-wall for the small-scale turbulence intensity that extends well in to the boundary layer. 

The large-scale variance, however, increases with increasing Reynolds number.  This seems to confirm that the value of $A^+$ should indeed be independent of wall-normal location (at least in the near-wall and logarithmic region). It should be noted that this value of $A^+$ is a conservative bound on the frequency. In time-scale, this corresponds to 200 wall-units. The physical significance of 200 wall-units as the cut-off time scale is obvious. The near-wall streaks in boundary layers/pipes/channels all last approximately 100 wall-based time-units (and is equivalent to 1000 wall-units length scale) and their features are very robust regardless of the type of flow. Therefore, the cut-off of 200 wall-units essentially captures the energy contained in these near-wall streaks that self-sustain themselves given a certain wall-shear-stress. 

This value of $A^+$ should be considered as high frequency bound above which spectral law-of-the-wall is expected to hold. It is possible that at increasing Reynolds numbers, spectral collapse is observed up to lower values of $f^+$ (i.e. lower frequencies or equivalently wavenumbers - which is indeed the case in figure \ref{fig1}) due to increasing scale separation between inner and outer scales. In fact, there is also evidence that a cut-off wavelength of 7000 to 10000 wall-units results in the collapse of small-scale turbulence statistics in the near-wall region (see \cite{Marusic2010a,Smits2011}). This length-scale is derived from frequency spectra in liaison with Taylor's local hypothesis. Ref. \cite{Mathis2009} used a cut-off wavelength of 7000 wall-units, which approximately corresponds to $f^+ \approx  0.001$, to separate the inner and outer scales to measure amplitude modulation. Moreover, recently,  Ref.\cite{Baars2017} and Ref.\cite{Marusic2017} showed that the spectral coherence between two probes (one located in the near-wall region and another in the outer region) at high Reynolds numbers reduced to zero at around 7000 wall-units, indicating a value of $A^+ = 0.005$ would also not have any coherence. Therefore, the proposed time-scale of 200 wall-units is a very conservative estimate for the cut-off frequency above which spectral similarity should hold. 

\begin{figure}
\begin{center}
\includegraphics[width=0.55\textwidth]{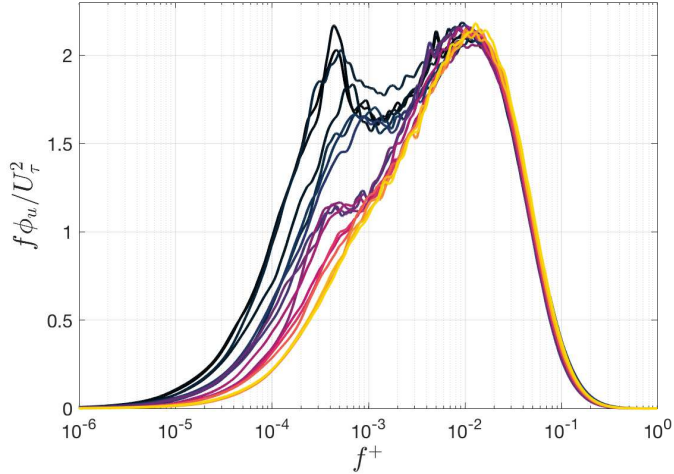}
\caption{Inner-normalised pre-multiplied spectra at $y^+ \approx 15$ for different levels of freestream turbulence. Lines are coloured from lightest to darkest in order of increasing turbulence intensity. The abscissa shows inner-normalised frequency $f^+$. Further details on the different turbulence intensities can be seen in \cite{Dogan2016,Dogan2017}.}
\label{fig3}
\end{center}
\end{figure}

The influence of outer region on the collapse of near-wall high-frequency spectra can be further examined by using data from experiments where the outer influence is artificially enhanced by the introduction of free-stream turbulence. Figure \ref{fig3} shows spectra obtained at $y^+ =15$ in an experiment that was designed to increase the strength of outer influence \cite{Dogan2016,Dogan2017}. The figure shows the premultiplied power-spectral density (normalised by $U_\tau$) versus inner normalised frequency. In these experiments, the outer influence was exacerbated by using freestream turbulence and this turbulence intensity would penetrate the boundary layer and alter the mean skin-friction characteristics as well as spectral properties \cite{Dogan2016}. However, when this altered skin-friction is taken in to account, the collapse of the spectra at $y^+ = 15$ over a wide range of external forcing is really clear (freestream turbulence levels up to 15\%). It can be seen that for low freestream turbulence intensity, the collapse is excellent till $f^+$ = 0.001. However, as the FST levels increases, the energetic outer scales penetrate down to the wall and the collapse is worse. Yet, the spectra appear to collapse up to $f^+$ = 0.005 regardless of the intensity of forcing in the outer layer.

\section{Conclusions}
The dimensional analysis approach of \cite{perry1977} and \cite{Zamalloa2014} to obtain spectral scaling is re-interpreted in the context of experimental observations in high Reynolds number turbulent boundary layers. A scaling relation for the high frequency regime of streamwise energy spectra (in frequency domain) in the near-wall region is proposed based on the hypothesis of law-of-the-wall for the small-scale turbulence intensity. The observations here are consistent with the observations in internal flows at lower Reynolds numbers \cite{Zamalloa2014}. In frequency domain, the spectra at similar wall-normal positions are found to collapse down to $f^+ \approx$0.005 across a whole range of Reynolds numbers. The corresponding small-small turbulence intensity also appears to collapse across all wall-normal positions and follows a logarithmic decay with wall-normal position farther from the wall, providing further support for law-of-the-wall for the small-scale turbulence intensity. These observations were further confirmed using experimental data where the boundary layer is under the influence of free-stream turbulence. 

\acknowledgements
We are grateful to Prof. Marusic and Prof. Hutchins for making the hot-wire data from HRNBLWT available. We thank Dr. Jason Hearst for providing Figure \ref{fig3}. We are thankful to Prof. Marusic for his comments on an earlier version of this manuscript.  We also acknowledge the financial support of the European Research Council (ERC Grant agreement No.\ 277472), and the Engineering and Physical Sciences Research Council of the United Kingdom (EPSRC Grant Ref.\ No.\ EP/L006383/1).

\section*{Data statement}
No new data were created during this study. 

%%%%%%%%%%%% BIBLIOGRAPHY %%%%%%%%%%%%%%%%%%%%%
\bibliographystyle{jfm}
\bibliography{law_of_the_wall}

\begin{thebibliography}{14}
\expandafter\ifx\csname natexlab\endcsname\relax\def\natexlab#1{#1}\fi
\def\au#1{#1} \def\ed#1{#1} \def\yr#1{#1}\def\at#1{#1}\def\jt#1{\textit{#1}}
  \def\bt#1{#1}\def\bvol#1{\textbf{#1}} \def\vol#1{#1} \def\pg#1{#1}
  \def\publ#1{#1}\def\arxiv#1{#1}\def\org#1{#1}\def\st#1{\textit{#1}}

\bibitem[Baars {\em et~al.\/}(2017)Baars, Hutchins \& Marusic]{Baars2017}
{\sc \au{Baars, WJ.}, \au{Hutchins, N} \& \au{Marusic, I}} \yr{2017}
  \at{Self-similarity of wall-attached turbulence in boundary layers}.
  \jt{Journal of Fluid Mechanics}  \bvol{823},  \pg{R2}.

\bibitem[Dogan {\em et~al.\/}(2016)Dogan, Hanson \&
  Ganapathisubramani]{Dogan2016}
{\sc \au{Dogan, E}, \au{Hanson, RE} \& \au{Ganapathisubramani, B}} \yr{2016}
  \at{Effects of large-scale freestream turbulence on turbulent boundary
  layers}.  \jt{Journal of Fluid Mechanics}  \bvol{802},  \pg{79--107}.

\bibitem[Dogan {\em et~al.\/}(2017)Dogan, Hearst \&
  Ganapathisubramani]{Dogan2017}
{\sc \au{Dogan, E}, \au{Hearst, RJ} \& \au{Ganapathisubramani, B}} \yr{2017}
  \at{Modelling high {R}eynolds number wall--turbulence interactions in
  laboratory experiments using large-scale free-stream turbulence}.
  \jt{Philosophical Transactions of the Royal Society A}  \bvol{375}~(2089),
  \pg{20160091}.

\bibitem[HUTCHINS {\em et~al.\/}(2011)HUTCHINS, MONTY, GANAPATHISUBRAMANI, NG
  \& MARUSIC]{Hutchins2011}
{\sc \au{HUTCHINS, N.}, \au{MONTY, J.~P.}, \au{GANAPATHISUBRAMANI, B.}, \au{NG,
  H. C.~H.} \& \au{MARUSIC, I.}} \yr{2011}  \at{Three-dimensional conditional
  structure of a high-reynolds-number turbulent boundary layer}.  \jt{Journal
  of Fluid Mechanics}  \bvol{673},  \pg{255--285}.

\bibitem[HUTCHINS {\em et~al.\/}(2009)HUTCHINS, NICKELS, MARUSIC \&
  CHONG]{Hutchins2009}
{\sc \au{HUTCHINS, N.}, \au{NICKELS, T.~B.}, \au{MARUSIC, I.} \& \au{CHONG,
  M.~S.}} \yr{2009}  \at{Hot-wire spatial resolution issues in wall-bounded
  turbulence}.  \jt{Journal of Fluid Mechanics}  \bvol{635},  \pg{103--136}.

\bibitem[Marusic {\em et~al.\/}(2017)Marusic, Baars \& Hutchins]{Marusic2017}
{\sc \au{Marusic, I}, \au{Baars, W~J.} \& \au{Hutchins, N}} \yr{2017}
  \at{Scaling of the streamwise turbulence intensity in the context of
  inner-outer interactions in wall turbulence}.  \jt{Phys. Rev. Fluids}
  \bvol{2},  \pg{100502}.

\bibitem[Marusic {\em et~al.\/}(2010)Marusic, Mathis \& Hutchins]{Marusic2010a}
{\sc \au{Marusic, I}, \au{Mathis, R} \& \au{Hutchins, N}} \yr{2010}  \at{High
  reynolds number effects in wall turbulence}.  \jt{International Journal of
  Heat and Fluid Flow}  \bvol{31}~(3),  \pg{418 -- 428}.

\bibitem[Mathis {\em et~al.\/}(2009)Mathis, Hutchins \& Marusic]{Mathis2009}
{\sc \au{Mathis, R}, \au{Hutchins, N} \& \au{Marusic, I}} \yr{2009}
  \at{Large-scale amplitude modulation of the small-scale structures in
  turbulent boundary layers}.  \jt{Journal of Fluid Mechanics}  \bvol{628},
  \pg{311--337}.

\bibitem[Perry \& Abell(1977)]{perry1977}
{\sc \au{Perry, AE} \& \au{Abell, CJ}} \yr{1977}  \at{Asymptotic similarity of
  turbulence structures in smooth- and rough-walled pipes}.  \jt{Journal of
  Fluid Mechanics}  \bvol{79}~(4),  \pg{785--799}.

\bibitem[Perry \& Maru{\v{s}}i{\'c}(1995)]{Perry1995}
{\sc \au{Perry, AE} \& \au{Maru{\v{s}}i{\'c}, Ivan}} \yr{1995}  \at{A wall-wake
  model for the turbulence structure of boundary layers. part 1. extension of
  the attached eddy hypothesis}.  \jt{Journal of Fluid Mechanics}  \bvol{298},
  \pg{361--388}.

\bibitem[Perry \& Chong(1982)]{Perry1982}
{\sc \au{Perry, A.~E.} \& \au{Chong, M.~S.}} \yr{1982}  \at{On the mechanism of
  wall turbulence}.  \jt{Journal of Fluid Mechanics}  \bvol{119},
  \pg{173--217}.

\bibitem[Perry {\em et~al.\/}(1986)Perry, Henbest \& Chong]{Perry1986}
{\sc \au{Perry, A.~E.}, \au{Henbest, S.} \& \au{Chong, M.~S.}} \yr{1986}  \at{A
  theoretical and experimental study of wall turbulence}.  \jt{Journal of Fluid
  Mechanics}  \bvol{165},  \pg{163--199}.

\bibitem[Smits {\em et~al.\/}(2011)Smits, McKeon \& Marusic]{Smits2011}
{\sc \au{Smits, AJ}, \au{McKeon, BJ} \& \au{Marusic, I}} \yr{2011}
  \at{High-€"reynolds number wall turbulence}.  \jt{Annual Review of Fluid
  Mechanics}  \bvol{43}~(1),  \pg{353--375}.

\bibitem[Zamalloa {\em et~al.\/}(2014)Zamalloa, Ng, Chakraborty \&
  Gioia]{Zamalloa2014}
{\sc \au{Zamalloa, CZ}, \au{Ng, HCH}, \au{Chakraborty, P} \& \au{Gioia, G}}
  \yr{2014}  \at{Spectral analogues of the law of the wall, the defect law and
  the log law}.  \jt{Journal of Fluid Mechanics}  \bvol{757},  \pg{498--513}.

\end{thebibliography}

\end{document}